\RequirePackage{ifpdf}
\documentclass[hyper,letterpaper]{JHEP3}
\pdfoutput=1
\usepackage{amsmath,amssymb,amsfonts,bm,amscd}
\usepackage{cite}
\usepackage{graphicx}
\usepackage{multirow}
\usepackage{verbatim}
\usepackage{appendix}
\usepackage{fancybox}
\usepackage{array}
\usepackage{url}
\usepackage{float}

\newcommand{\Z}{{\mathbb Z}}

\newcommand{\C}{{\mathbb C}}
\newcommand{\Q}{{\mathbb Q}}

\def\Tr{{\rm Tr \,}}

\def\CC{{\mathcal C}}
\def\CE{{\mathcal E}}

\def\CH{{\mathcal H}}
\def\CI{{\mathcal I}}

\def\CL{{\mathcal L}}

\def\CN{{\mathcal N}}

\def\be{\begin{equation}}
\def\ee{\end{equation}}
\def\bea{\begin{eqnarray}}
\def\eea{\end{eqnarray}}

\def\cn{{\mathtt p}}
\def\fn{{\mathtt q}}

\renewcommand{\d}{\partial}

\def\sym{\text{Sym}^2\Box}
\def\afund{\bar\Box}

\title{Exact solutions of (0,2) Landau-Ginzburg models}

\author{Abhijit Gadde, Pavel Putrov
\\
School of Natural Sciences, Institute for Advanced Study, Princeton, NJ 08540, USA
}

\abstract{
In this paper we study the low energy physics of Landau-Ginzburg models with ${\cal N}=(0,2)$ supersymmetry. We exhibit a number of classes of relatively simple LG models where the conformal field theory at the low energy fixed point can be explicitly identified. One interesting class of fixed points can be thought  of as ``heterotic" minimal models.  Other examples include ${\cal N}=(0,2)$ renormalization group flows that end up at  ${\cal N}=(2,2)$ minimal models and models with non-abelian symmetry.}

\begin{document}

\section{Introduction}
Landau-Ginzburg theories with ${\cal N}=(0,2)$ supersymmetry describe certain phases of $(0,2)$ supersymmetric gauge theories. Their usefulness towards understanding the $\CN=(0,2)$ Calabi-Yau sigma model is well known \cite{Distler:1993mk, Blumenhagen:1995ew}. Landau-Ginzburg orbifolds and their elliptic genera have been studied in \cite{Kawai:1994np}. Recently, a connection between geometry and topology of four manifolds and $\CN=(0,2)$ supersymmetric theories have been established \cite{Gadde:2013sca}. This led the authors to study the dynamics $\CN=(0,2)$ gauge theories \cite{Gadde:2013lxa, Gadde:2015kda} and determine their low energy fixed point theory \cite{Gadde:2014ppa}. 
In this paper, we study the LG models on their own right focusing on identifying their low energy physics exactly. The philosophy of the paper is similar to that of \cite{Gadde:2014ppa} where the low energy physics of a class of (0,2) supersymmetric gauge theories was identified using arguments involving 't Hooft anomaly matching, c-extremization and modular invariance of the partition function on the torus. 

By definition, a Landau-Ginzburg models has discreet vacua. This gives rise to a normalizable vacuum state in the quantum theory. A conformal field theory with normalizable vacuum enjoys a  ``state-operator correspondence". Using this correspondence it is straightforward to argue that a global symmetry of the microscopic theory is enhanced to either a holomorphic or an anti-holomorphic chiral symmetry. This includes the supersymmetry i.e. the two supersymmetries of the microscopic $(0,2)$ theory are promoted to the \emph{chiral} supercurrents. Along with the chiral stress-tensor and R-current, the supercurrents form the $\CN=2$ super-Virasoro algebra.
Requiring the vanishing of the commutator between R-symmetry and other abelian symmetries, the central charge of the super-Virasoro algebra can be determined. This prescription is known as  c-extremization \cite{Benini:2013cda}. This is one of the important tools we use. It rests on the assumption that there are no new abelian symmetries in the infra red. When this assumption is not valid, a modification of this procedure \cite{Bertolini:2014ela} is applied as we demonstrate in some of our examples.

On these general grounds, we expect the $(0,2)$ Landau-Ginzburg theory to flow to a heterotic  SCFT with the left-moving  spectrum governed by Virasoro symmetry and right-moving spectrum governed by $\CN=2$ super-Virasoro symmetry. The partition function of the theory on the torus i.e.
\be
Z={\rm Tr}\, q^{L_0} \bar q^{\bar L_0}
\ee
is invariant under the modular transformations\footnote{We will consider the partition function with anti-periodic boundary conditions for the fermions along both cycles of the torus. In this sector the partition function is expected to have invariance under the subgroup $\Gamma_0(2)$.} $\tau\to (a\tau+b)/(c\tau+d)$ for $a,b,c,d \in {\mathbb Z}$ and $q=e^{2\pi i \tau}$. This condition puts a strong constraint on the spectrum apart from the symmetries. In the class of examples we study, the above considerations turn out to be strong enough to determine the low energy theory completely.


\section{Search for solvable LG models}
\label{sec:search}

Before we start the search for solvable Landau-Ginzburg model, a quick introduction to their Lagrangian is in order. A $(0,2)$ Landau-Ginzburg model is constructed using $\cn$ chiral superfields $\Phi_i$s and $\fn$ Fermi superfields $\Psi_a$. The chiral multiplet consists of a complex scalar $\phi$ and a complex right-moving fermion $\lambda$ and the Fermi multiplet consists of a single complex left-moving fermion $\psi$. 
The supersymmetry allows for two types of interaction terms, the $J$-type and the $E$-type. The $J$-type interaction is analogous to the superpotential term. Most compactly, it is presented as the integral over half the superspace
\be
\int d\theta^+ \sum_a \, \Psi_a J_a(\Phi_i)+\, {\rm c.c.}
\ee
where $J_a$ are holomorphic functions of $\Phi_i$. For brevity, we will drop c.c. from now on.
The $E$-type interaction is induced somewhat unconventionally as supersymmetry variation of the Fermi field i.e. by requiring $\bar D_+ \Psi_a=E_a(\Phi_i)$ instead of $\bar D_+ \Psi_a=0$. The $\CN=(0,2)$ supersymmetry requires $\sum_a E_aJ_a=0$. 
In terms of component fields, these interactions get spelled out as follows
\be
\CL= \ldots - \sum_a \Big( |J_a|^2+|E_a|^2 \Big)- \sum_a \sum_i \Big(\psi_a \bar \lambda_i \frac{\partial J_a}{\partial \phi_i}+ \bar \psi_a \lambda_i \frac{\partial E_a}{\partial \phi_i}+{\rm c.c.}\Big).
\ee
The $E_a$ and the $J_a$-type interactions are interchanged by conjugating the Fermi multiplet $\Psi_a$. This means the action of supersymmetry on $\bar \Psi$ is given by $\bar D_+ \bar \Psi_a= J_a$. In this paper, we will set all the $E$-terms to zero. Written compactly, the tree level supercharge is
\be\label{susyaction}
\bar D_+= J_a \frac{\delta}{\delta \bar \Psi_a}. 
\ee
This gives a straightforward way of computing the tree level cohomology of the theory. Interestingly, the tree level cohomology it is known to be quantum mechanically exact \cite{Dedushenko:2015opz}. As a vector space it gives a BPS sector of the Hilbert space $\CH$ of the low energy theory on the circle:
\begin{equation}
	\CH_\text{BPS}=\left.\CH\right|_{\bar{L}_0+\bar{J}_0/2=0}
\end{equation}
where, as usual, $\bar{L}_0$ is the anti-holomorphic conformal dimension and $\bar{J}_0$ is the infra-red R-charge. 
The translations in the anti-holomorphic direction are cohomologically exact so $\CH_\text{BPS}$ can be thought of as the Hilbert space of a holomorphic conformal field theory. 
This holomorphic theory is essentially the topologically half-twisted  theory \cite{Melnikov:2007xi}. 

In a favorable situations one can consider a particular sector $\CH_\text{Top}\subset\CH_\text{BPS}$, the \textit{topological heterotic ring} of the (0,2) 2d theory \cite{Adams:2005tc}. It is a finite dimensional subspace generated by the elements saturating the following bound:
\begin{equation}
 L_0\geq \mathfrak{q}/2
\label{Top-bound}
\end{equation}
where $\mathfrak{q}$ is charge of a certain left-moving $U(1)_\text{L}$ symmetry. This special left-moving $U(1)_\text{L}$ symmetry is uniquely fixed by condition that it has the same charges for the chiral fields $\Phi_i$ as the right-moving R-symmetry $\bar J_0$. This also means $\mathfrak{q} = \bar J_0-1$ for Fermi fields.
The hierarchy of vector spaces is as follows.
\begin{equation}
 \CH_\text{Top}=\CH_\text{BPS}|_{L_0=\mathfrak{q}/2}=
\CH|_{\bar{L}_0+\bar{J}_0/2=0,\;L_0=\mathfrak{q}/2}
\end{equation} 
Morally, the topological heterotic ring $\CH_\text{Top}$ can be understood as the zero-mode sub-sector of the holomorphic CFT $\CH_\text{BPS}$ where one first performs a further ``topological half-twist" by $U(1)_\text{L}$. Strictly speaking this is not a topological twist as $U(1)_\text{L}$ is not an R-symmetry.

 As was shown in \cite{Melnikov:2009nh} the topological heterotic ring of $(0,2)$ Landau-Ginzburg models can be obtained in terms of  Koszul homology of the complex
\begin{equation}
\CC=\qquad0\stackrel{d}{\longrightarrow}\wedge^\fn \CE
\stackrel{d}{\longrightarrow}
\ldots 
\stackrel{d}{\longrightarrow}\wedge^1 \CE
\stackrel{d}{\longrightarrow}\wedge^0 \CE
\stackrel{d}{\longrightarrow}0
\end{equation}
where
\begin{equation} \label{koszul}
\CE=\text{Span}_\C\{\bar\Psi_a\}_{a=1}^\fn\otimes \C[\Phi_i]
\cong
\C[\Phi_i]^\fn
\end{equation}
and the differential is given by the interior derivative
\begin{equation}
	d=\imath_J,\qquad J=\sum_{a=1}^\fn \Psi_a\,J_a\;\in\;\CE^*.
\end{equation}
This follows  from \eqref{susyaction}.
The Koszul homology,
\begin{equation}
H_n(\CC,d)=\mathrm{Ker}\,d|_{\wedge^n\CE}\,/\,d(\wedge^{n-1}\CE),
\end{equation}
gives the topological heterotic ring in terms of microscopic fields. That is,
\begin{equation}
  H_*(\CC,d)\cong \CH_\text{Top}
\end{equation}  
Its calculation in straightforward and is much easier than calculation of the full infinite dimensional $\bar{Q}$-cohomology $\CH_\text{BPS}$. Yet it contains a non-trivial information about the low energy superconformal field theory and can be used to check the proposed IR solutions of Landau-Ginzburg models.

The superconformal index $\CI$ of an $\CN=(0,2)$ LG theory in the NSNS sector is computed even more easily. It is the super-trace over the $\bar Q$-cohomology $\CH_\text{BPS}$.
 Let the charge of chiral fields and Fermi fields be $q_{(\ell)}^{\Phi_i}$ and $q_{(\ell)}^{\Psi_i}$ respectively under the abelian symmetry $U(1)^{(\ell)}$. Let $u_{(\ell)}$ be the chemical potential for this symmetry. The superconformal index takes a compact form,
\be
\CI= \frac{\prod_{a=1}^\fn \, \theta(q^{\frac{1+r_{\Psi_a}}{2}} \, e^{\sum_\ell u_{(\ell)} \,{q_{(\ell)}^{\Psi_a}}};q)}{\prod_{i=1}^\cn\, \theta(q^{\frac{r_{\Phi_i}}{2}}\,e^{\sum_\ell u_{(\ell)} \,{q_{(\ell)}^{\Phi_i}}};q)}
\ee
where $r_{\Phi_i}$ and $r_{\Psi_a}$ are the R-charge of chiral fields and Fermi fields  respectively.  The superconformal index can be projected to the trace over the topological heterotic ring by setting $L_0 = \mathfrak{q}/2$ where $\mathfrak{q}$ is equal to $r_\Phi$ for chiral fields and $r_{\Psi}-1$ for the Fermi fields. 
As a result, the trace over topological heterotic ring or equivalently the equivariant Euler characteristic of Koszul homology is  
\be\label{eulerchar}
\CI \,\, \mathop{\longrightarrow}_{q\rightarrow 0}^{\text{twist}} \,\,\, \chi  = \frac{\prod_{a=1}^{\fn}\Big(1-e^{-\sum_\ell u_{(\ell)} \,{q_{(\ell)}^{\Psi_a}}}\Big)}{\prod_{i=1}^\cn \Big(1-e^{\sum_\ell u_{(\ell)} \,{q_{(\ell)}^{\Phi_i}}}\Big)}.
\ee

Finally it is useful to note that an $\CN=(2,2)$ superfield $\hat \Phi$ decomposes into a $(0,2)$ chiral superfield $\Phi$ and a Fermi superfield $\Psi$. The superpotential $W(\hat \Phi_i)$ results in the $J$-term interaction $J_i=\partial W(\Phi)/\partial \Phi_i$. So whenever the $J$-term is a gradient, the $(0,2)$ Lagrangian is actually $(2,2)$ supersymmetric. In this case the topological heterotic ring coincides with the usual chiral ring:
\begin{equation}
	\CH_\text{Top}=\C[\Phi_i]/\{\d W(\Phi_i)/\d \Phi_i\}_i\,.
\end{equation}

We are now ready to begin the search for solvable Landau-Ginzburg theories. 
Each Fermi multiplet yields a relation $J_a=0$ on the moduli space of chiral superfield. 
In order for the $\CN=(0,2)$ Lagrangian to have discreet vacua, the number of Fermi multiplets $\fn$ should be greater than or equal to the number of chiral multiplets $\cn$. We start our search for solvable LG model with $\fn\geq \cn$ at  $\cn=1,\fn=1$. In this case, the only $J$-term interaction is $J=\Phi^n$ for some $n$. This is a total derivative. The Lagrangian is that of a $(2,2)$ Landau-Ginzburg model with $W=\hat \Phi^{n+1}$. This model has the central charge $c_L=c_R=3(n-1)/(n+1)$ and flows to $(n-1)$-th $\CN=2$ supersymmetric minimal model \cite{Witten:1993jg}. The first non-trivial example is a theory with $\cn=1$, $\fn=2$.

\subsection{$\cn=1$, $\fn=2$}
The only $J$-term interaction that is consistent with R charge assignment is
\be\label{p1q2}
\int d\theta^+ \, \Psi_1 \Phi^n +\Psi_2 \Phi^m \qquad \qquad m,n \in {\mathbb Z}_+.
\ee
The right-moving central charge of this model can be determined by $c$-extremization. Assigning R-charge $r$ to chiral superfield $\Phi$,
\be
c_R=3\,{\rm Tr}\gamma^3 R^2= 3\Big( (r-1)^2-(1-n r)^2-(1-m r)^2\Big).
\ee
Extremizing this function, we get 
\be
r=\frac{m+n-1}{m^2+n^2-1} \quad \Rightarrow \quad c_R=\frac{6(m-1)(n-1)}{m^2+n^2-1}.
\ee
Interestingly $c_R \leq 3$ with inequality saturating at $m,n\to \infty$. This implies that the central charge has to be equal to that of a $\CN=2$ minimal model i.e. $3k/(k+2)$ for some $k\in {\mathbb Z}_+$. But surprisingly that is not the case. This presents an interesting puzzle. The solution is in realizing that the $c$-extremization procedure is not valid if there is an enhanced symmetry in the infrared. This is indeed the case here. Let us elaborate.

First set $m=n$, only the linear combination $\Psi_1+\Psi_2$ couples to the chiral multiplet and the other combination is free. In this case we get an $\CN=(2,2)$ minimal model, as before tensored with a complex left-moving fermion. For $m=n+1$, the $c$-extremization yields $c_R=3(n-1)/(n+1)$. This is of the form $3k/(k+2)$ for $k=n-1$. We conclude that the right-moving part of the low energy CFT is the $\CN=2$ minimal model with $k=n-1$. Precisely in this case, the $R$-charge of the Fermi multiplet $\Psi_2$ is $0$ hence, $\bar L_0=0$. Using the fact that the spin $L_0-\bar L_0$ is $\frac12$ we get $L_0=\frac12$. This is the unitarity bound where the complex left-moving fermions on $\Psi_2$ become free. At these values of $m$ and $n$ there is an extra $U(1)$ symmetry that rotates only $\Psi_2$. 

For $m \geq n+1$, the naive application of c-extremization yields unitarity bound violating left-moving fermion $\Psi_2$. This phenomenon is reminiscent of the one that occurs in four dimensional supersymmetry QCD \cite{Kutasov:2003iy}. In the context of $2d$ theories, it was studied in \cite{Bertolini:2014ela}.
Following them, we take this to signal the decoupling of $\Psi_2$ from the interacting theory.  It is accounted for by taking the correct contribution of $\Psi_2$ to the right-moving central charge
\be
c_R^{\rm new}=c_R^{\rm old}+(1-m r)^2-0.
\ee
Extremizing with respect to $r$, we get $c_R=3(n-1)/(n+1)$, same as before. This is of the form $3k/(k+2)$ for $k=n-1$. The fermion $\Psi_2$ is free and so we can ignore the corresponding $J$-term interaction. This means the low energy theory consists of $(n-1)$-th $\CN=(2,2)$ minimal model tensored with a free complex left-moving fermion. 

We can verify this conclusion by computing the topological heterotic ring of the theory using \eqref{koszul}. 
\bea
&& H_2(\CC,d)=0,\\
&& H_1(\CC,d)\cong {\rm Span}_{\mathbb C} \{\bar \Psi_1 \Phi^i -\bar \Psi_2 \Phi^{n+i-m}\}_{i=0}^{n-1},\\
&& H_0(\CC,d)\cong {\rm Span}_{\mathbb C} \{ \Phi^i \}_{i=0}^{n-1}.
\eea
As a vector space the cohomology above agrees with that of the proposed low energy theory. The $(n-1)$-th $\CN=2$ minimal model has $H_0$ of dimension $n$ which is tensored with the free left-moving fermion $\bar \Psi_2$ to generate $H_1$. 
This analysis can be straightforwardly generalized to $\cn=1$ and $\fn >2$.

\subsection{$\cn=2$, $\fn=3$}
With two chiral superfields and three Fermi superfields\footnote{The analogous case with $\cn=2$, $\fn=2$,  which could be expected to be simpler actually turns out to be more subtle. We consider it in the next section.} we can write an interesting model that  is solvable. 
\begin{equation}
 \int d\theta^+\,\left(\Psi_1\Phi_1^m+\Psi_2\Phi_2^n+\Psi_3\Phi_1\Phi_2\right) \qquad\qquad m,n\in {\mathbb Z}_+
 \label{second-potential}
\end{equation} 
We compute the right-moving central charge using $c$-extremization and the left-moving central charge can be determined from the gravitational anomaly \eqref{gravan}, $c_R-c_L=-1$.
Taking the $r$-charge of $\Phi_1$ and $\Phi_2$ to be $r_{\Phi_1}$ and $r_{\Phi_2}$, the trial central charge is
\be
c_R=3\Big((r_{\Phi_1}-1)^2+(r_{\Phi_2}-1)^2-(1-m\, r_{\Phi_1} )^2-(1-n\, r_{\Phi_2} )^2-(1- r_{\Phi_1}-r_{\Phi_2} )^2\Big).
\ee
Extremizing with respect to $r_{\Phi_1}$ and $r_{\Phi_2}$
\be \label{r-charges}
r_{\Phi_1}=\frac{n}{mn+1},\, r_{\Phi_2}=\frac{m}{mn+1} \qquad \Rightarrow \qquad c_R=3\frac{mn-1}{mn+1},\,c_L=2\frac{2mn-1}{mn+1}.
\ee
For no value of $m,n$ does the $r$-charge of Fermi fields becomes negative. This justifies the $c$-extremization prescription. Interestingly, in this case we also find $c_R<3$. This means that the right-moving part of the low energy CFT is a chiral of the $(mn-1)$-th $\CN=2$ minimal model. To identify the left-moving part we note that for general values of $m,n$ the theory has $U(1)\times U(1)$ symmetry.  Let us denote $U(1)^{(i)}$ the symmetry with respect to which $\Phi_j$ has charge $\delta_{ij}$. The $U(1)^{(1)}\times U(1)^{(2)}$ charges of $\Psi_{1,2,3}$ are then $(-m,0)$, $(0,-n)$ and $(-1,- 1)$ respectively. The symmetrized anomaly matrix for these symmetries is
\be
Q=\left(
\begin{array}{cc}
m^2 & 1\\
1 & n^2
\end{array}
\right)
\ee
The left-moving CFT contains $(U(1)\times U(1))_Q$ chiral symmetry. These symmetries contribute $2$ to the left-moving central charge. The remaining piece should have $c=2(mn-2)/(mn+1)$. We propose that this left-over central charge is contributed by the $SU(2)/U(1)$ coset. The low energy theory takes the form
\be
\Big(\frac{SU(2)_{mn-1}}{U(1)_{2(mn-1)}} \times (U(1)\times U(1))_Q \Big) \otimes \overline{\Big( \frac{SU(2)_{mn-1} \times SO(2)_1}{U(1)_{2(mn+1)}} \Big)}.
\label{CFTp2q3}
\ee
Here we have use the coset representation of the $\CN=2$ minimal model. 

The existence of modular invariant partition function is guaranteed by  the following equivalence of quadratic forms over rational numbers (\textit{cf.} \cite{gannon1991gluing,gannon1997u})
\begin{equation}
	Q\oplus 2(mn+1)\;\stackrel{\Q}{\sim}\; \text{Id}_{\Z^2}\oplus 2(mn-1).
	\label{Qequiv}
\end{equation}
Namely, that there exists $\text{GL}(3,\Q)$ transformation $(u,v,w)\rightarrow (x,y,z)$ such that
\begin{equation}
	m^2u^2+n^2v^2+2uv+2(mn+1)w^2\;\equiv\; x^2+y^2+2(mn-1)z^2.
\end{equation}
The equivalence (\ref{Qequiv}) then can be shown by the following transformation:
\begin{equation}
	\begin{array}{rl}
		x = & mu+v/m, \\
		y = & v(mn-1)/m+2w, \\
		z = & v/m-w.
	\end{array}
	\label{Qtransform}
\end{equation}

Let us show how one can explicitly construct a modular invariant partition function for (\ref{CFTp2q3}) using the explicit transformation (\ref{Qtransform}). Namely, we want to find coefficients $C$ appearing in the decomposition of the partition function into WZW characters\footnote{Note that the explicit expression for characters depends on the choice of periodic/anti-periodic boundary conditions on the torus.}:
\begin{equation}
	Z=\sum_{\alpha,\beta,\lambda,\nu} C_{\nu,\bar{\beta},\bar\lambda}
	\,\chi_{\alpha;\bar{\nu}}^{SU(2)_{mn-1}/U(1)_{2(mn-1)}}
	\chi_{\lambda}^{U(1)^2_Q}\cdot 
	\bar{\chi}^{SU(2)_{mn-1}\times SO(2)_1/U(1)_{2(mn+1)}}_{\bar{\alpha};\beta}\
	\label{Z2explicit}
\end{equation}
where a bar over an index means that the corresponding quantity transforms in the conjugate representation of the modular group. Denote by $R_{\hat{\mathfrak{g}}_k}$ a linear finite dimensional representation of the modular group for which the basis is formed by the characters $\chi^{\hat{\mathfrak{g}}_k}_\mu(q)$ of the affine algebra $\hat{\mathfrak{g}}$ at level $k$.
In order for expression (\ref{Z2explicit}) to be modular invariant the coefficients $C$ should form an invariant  tensor of the type
\begin{equation}
	C\;\;\in \;\;R_{U(1)_{2(mn-1)}}\otimes  \bar{R}_{U(1)_{2(mn+1)}} \otimes \bar{R}_{U(1)_Q^2}.
	\label{Ctensor}
\end{equation}
It can be constructed explicitly in two steps as follows. 
 
 First let us remind that the space $R_{\hat{\mathfrak{g}}_k}$, spanned by the characters of $\mathfrak{g}_k$, can be interpreted as the space of holomorphic sections of a certain line bundle on $(T_\tau^2)^{\mathrm{rank}\,{\mathfrak{g}}}$ where $T^2_\tau$ is a 2-torus with modulus $\tau$. The transition functions of the bundle are determined by the choice of the level $k$, or, equivalently, by the corresponding anomaly quadratic form. When the section is explicitly represented as a function of chemical potentials $z_i\in T^2_\tau,\;i=1\ldots \mathrm{rank}\,{\mathfrak{g}}$, periodic with respect to $z_i\rightarrow z_i+2\pi i$, the quadratic form determines its transformation properties under the shifts $z_i\rightarrow z_i+2\pi i\tau$.
 
 From this point of view it follows that there should be the following decomposition of $U(1)_{2(mn-1)}\times U(1)_1\times U(1)_1$ characters into $U(1)^2_{m^2Q}\times U(1)_{2m^2(mn+1)}$ characters\footnote{Note that $u,v,w$, as well as $x,y,z$ denote chemical potentials, not fugacities.}:
\begin{multline}
	\chi_\nu^{U(1)_{2(mn-1)}}(mz)\,\chi^{U(1)_1}(mx)\,\chi^{U(1)_1}(my)=\\
	\sum_{\lambda',\beta'}A_{\nu,\bar{\lambda'},\bar{\beta'}}\,
	\chi^{U(1)^2_{m^2Q}}_{\lambda'}(u,v)\,\chi^{U(1)_{2m^2(mn+1)}}_{\beta'}(w)
\end{multline}
where we used the fact that when both sides of (\ref{Qtransform}) are multiplied by $m$ all coefficients become integers. Such decomposition provides us with an invariant tensor
\begin{equation}
	A\;\;\in \;\;R_{U(1)_{2(mn-1)}}\otimes \bar{R}_{U(1)_{m^2Q}^2}\otimes  \bar{R}_{U(1)_{2m^2(mn+1)}}. 
\end{equation}
The same argument tells us that there also should be decomposition
\begin{multline}
	\bar\chi_{\bar\lambda}^{U(1)^2_{Q}}(mu,mv)\,\bar\chi_{\bar\beta}^{U(1)_{2(mn+1)}}(mw)=
	\\
	\sum_{\lambda',\beta'}B_{\bar\lambda,\bar\beta,\lambda',{\beta'}}\,
	\bar\chi^{U(1)^2_{m^2Q}}_{\bar{\lambda'}}(u,v)\,
	\bar\chi^{U(1)_{2m^2(mn+1)}}_{\bar{\beta'}}(w)
\end{multline}
which gives us an invariant tensor
\begin{equation}
	B\;\;\in \;\;\bar{R}_{U(1)_{Q}^2}\otimes  \bar{R}_{U(1)_{2(mn+1)}} \otimes {R}_{U(1)_{m^2Q}^2}\otimes  {R}_{U(1)_{2m^2(mn+1)}}.
\end{equation}
Then the invariant tensor (\ref{Ctensor}) can be obtained by pairing of $A$ and $B$. Namely,
\begin{equation}
	C_{\nu,\bar{\beta},\bar{\lambda}}=\sum_{\lambda',\beta'}
	A_{\nu,\bar{\lambda'},\bar{\beta'}}
	B_{\bar\lambda,\bar\beta,\lambda',{\beta'}}.
\end{equation}
Although this procedure to obtain coefficients $C$ seems to break $m\leftrightarrow n$ symmetry, the final result (up to unimportant overall integer factor) still respects it. The sum (\ref{Z2explicit}) can be explicitly written as follows:
\begin{multline}
	Z=\sum_{\alpha=0}^{mn-1}\sum_{\nu\,\in\,\Z_{2(mn-1)}}\sum_{a\,\in\, \Z_{mn+1}}
	\,\chi_{\alpha;\bar{\nu}}^{SU(2)_{mn-1}/U(1)_{2(mn-1)}}
	\chi_{(ma,n(a+\nu))}^{U(1)^2_Q} \\
	\cdot\bar{\chi}^{SU(2)_{mn-1}\times SO(2)_1/U(1)_{2(mn+1)}}_{\bar{\alpha};2a+\nu}\
	\label{Z2veryexplicit}
\end{multline}
where $(ma,n(a+\nu))\in \mathrm{Coker}\,Q\equiv\Z^2/Q\Z^2$ if we treat $Q$ as an operator $Q:\Z^2\rightarrow \Z^2$. Note that the symmetry under exchange $m\leftrightarrow n$ can be seen via the following change of summation indices: $a=a'+\nu'$, $\nu=-\nu'$ and taking into account the fact that the $SU(2)/U(1)$ characters are invariant under $\bar\nu \leftrightarrow -\bar\nu$.

Of course, the formula (\ref{Z2veryexplicit}) for the partition function directly lifts to the following decomposition of the Hilbert space of the IR CFT on a circle:
\begin{multline}
	\CH=\bigoplus_{\alpha=0}^{mn-1}\bigoplus_{\nu\,\in\,\Z_{2(mn-1)}}\bigoplus_{\;\;a\,\in\, \Z_{mn+1}}
	\,\CH_{\alpha;\bar{\nu}}^{SU(2)_{mn-1}/U(1)_{2(mn-1)}}
	\otimes \CH_{(ma,n(a+\nu))}^{U(1)^2_Q} \\
	\otimes \bar{\CH}^{SU(2)_{mn-1}\times SO(2)_1/U(1)_{2(mn+1)}}_{\bar{\alpha};2a+\nu}\
	\label{H2veryexplicit}
\end{multline}
where $\CH_*^V$  and $\bar{\CH}_*^V$ denote the modules of holomorphic and anti-holomorphic vertex operator algebra (VOA) $V$ respectively.

For example consider the simplest case, $m=2$, $n=1$ in the NS sector. In the right-moving sector we then have $c_\text{R}=1$ \emph{i.e.} $\CN=2$ minimal model with 3 primaries:
\begin{equation}
	\begin{array}{llll}
		\alpha=0, \;\;& 2a+\nu=0 \mod 6,\;\; & \bar{L}_0=0, \;\;& \bar{J}_0=0, \\
		\alpha=1, & 2a+\nu=\pm 1 \mod 6,\;\;\; & \bar{L}_0=1/6, & \bar{J}_0=\pm 1/3,
	\end{array}
	\label{c1primaries}
\end{equation}
where $\bar{L}_0$ and $\bar{J}_0$ denote conformal dimensions and R-charges respectively. The $SU(2)/U(1)$ characters in the left-moving sector are 1 when $(\alpha,\nu)=(0,0)$ or $(\alpha,\nu)=(1,1)$ and zero otherwise. It follows that the sum (\ref{Z2veryexplicit}) has 3 non-zero terms corresponding to 3 primaries in (\ref{c1primaries}):
\begin{equation}
	\begin{array}{lll}
		\alpha=0, \;\;& \nu=0 \mod 2,\;\; & a=0 \mod 3,  \\
		\alpha=1, \;\;& \nu=1 \mod 2,\;\; & a=0 \mod 3,  \\
		\alpha=1, \;\;& \nu=1 \mod 2,\;\; & a=-1 \mod 3.  \\
	\end{array}
\end{equation}
The superconformal index picks only BPS primaries with $\bar{L}_0+\bar{J}_0/2=0$. This agrees with the UV index calculation\footnote{Note that the last theta-functions in the numerator and denominator cancel each other.}:
\begin{equation}
	\frac{\theta(q^{1/2}e^{-x-y})\theta(q^{2/3}e^{-2x})\theta(q^{2/3}e^{-y})}
	{\theta(q^{1/6}e^{x})\theta(q^{1/3}e^y)}=
	\chi_{(0,0)}^{U(1)^2_Q}(x,y)-\chi_{(-2,0)}^{U(1)^2_Q}(x,y).
\end{equation}

\subsubsection{Comparison with Koszul homology}

\begin{figure}[h]
\centering
\includegraphics[scale=1.7]{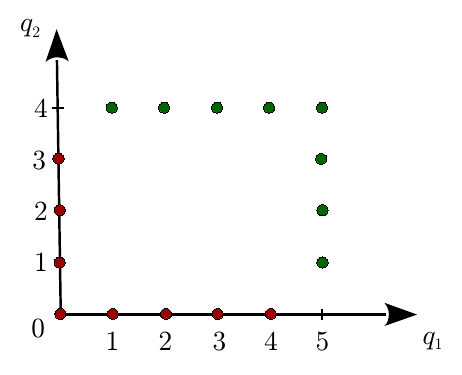}
\caption{The plane of $U(1)^{(1)}\times U(1)^{(2)}$ charges in the case $m=5$, $n=4$. The colored dots are in one-to-one correspondence with the generators of $H_*$. The red (green) dots correspond to the generators of $H_0$ ($H_1$).}
\label{fig:THRing}
\end{figure}

As an independent check of the validity of the IR solution let us show that it is consistent with the UV calculation of the topological heterotic ring (see section \ref{sec:search}). For the model with superpotential (\ref{second-potential}) the corresponding Koszul complex is the following:
\begin{equation}
\CC=\qquad0\stackrel{d}{\longrightarrow}\wedge^3 \CE
\stackrel{d}{\longrightarrow}\wedge^2 \CE
\stackrel{d}{\longrightarrow}\wedge^1 \CE
\stackrel{d}{\longrightarrow}\wedge^0 \CE
\stackrel{d}{\longrightarrow}0
\end{equation}
where
\begin{equation}
\CE=\text{Span}_\C\{\bar\Psi_i\}_{i=1}^3\otimes \C[\Phi_1,\Phi_2]
\cong
\C[\Phi_1,\Phi_2]^3.
\end{equation}
and the differential is given by the interior derivative:
\begin{equation}
d=\imath_J,\qquad J=\Psi_1\Phi_1^m+\Psi_2\Phi_2^n+\Psi_3\Phi_1\Phi_2\in \CE^*.
\end{equation}
It is easy to see that:
\begin{equation}
 \begin{array}{ccl}
   H_3(\CC,d) & = & 0,\\
   H_2(\CC,d) & = & 0,\\
   H_1(\CC,d) & \cong & \mathrm{Span}_\C\{\bar\Psi_3\Phi_1^{m-1}\Phi_2^b-\bar\Psi_1\Phi_2^{b+1} \}_{b=0}^{n-1} \oplus 
     \mathrm{Span}_\C\{\bar\Psi_3\Phi_2^{n-1}\Phi_1^a-\bar\Psi_2\Phi_1^{a+1} \}_{a=0}^{m-2},\\
   H_0(\CC,d)&\cong & \mathrm{Span}_\C\{\Phi_1^a \}_{a=0}^{m-1} \oplus 
        \mathrm{Span}_\C\{\Phi_2^b \}_{b=1}^{n-1},\\
 \end{array}
 \label{Koszul-explicit}
\end{equation}
where the generating elements in the right hand side can be understood as representatives of $H_*$ from $\CC$. Obviously, $d$ commutes with the generators of $U(1)\times U(1)$ flavor symmetry, so that $H_*$ is equipped with the corresponding $\Z^2$ grading. The generators in (\ref{Koszul-explicit}) have well defined $U(1)\times U(1)$ charges. Their spectrum is depicted in Figure \ref{fig:THRing}.

In the IR the BPS spectrum of the proposed solution reads 
\begin{equation}
	\CH_\text{BPS}\equiv \CH|_{\bar{L}_0+\bar{J}_0/2=0}=\bigoplus_{\alpha=0}^{mn-1}
\bigoplus_{\scriptsize\begin{array}{c} \nu\,\in\,\Z_{2(mn-1)}\\ a\,\in\, \Z_{mn+1} \\ 2a+\nu =-\alpha \mod 2(mn+1) \end{array}}
	\CH_{\alpha;\bar{\nu}}^{SU(2)_{mn-1}/U(1)_{2(mn-1)}}
	\otimes \CH_{(ma,n(a+\nu))}^{U(1)^2_Q} 
	\label{H2BPS}
\end{equation}
and has structure of a holomorphic $\Z_2$ graded\footnote{$\Z_2$ grading on $\CH_\text{BPS}=\CH|_{\bar{L}_0+\bar{J}_0/2=0}$ descends from $\Z_2$ (Fermion number) grading $2(L_0-\bar{L_0})\mod 2$ on the total Hilbert space $\CH$.}  CFT.
The topological heterotic ring forms its finite dimensional subspace
\begin{equation}
 \CH_\text{Top}=\CH_\text{BPS}|_{L_0=\mathfrak{q}/2}=
\CH|_{\bar{L}_0+\bar{J}_0/2=0,\;L_0=\mathfrak{q}/2}
\end{equation} 
with
\begin{equation}
 \mathfrak{q}=\frac{nq_1+mq_2}{mn+1}
\label{U1L-charge}
\end{equation}
where $q_\ell$ is $U(1)^{(\ell)}$ charge. This combination follows from the fact that $\mathfrak{q}$ defined above is equal to the R-charge for the chiral superfields \eqref{r-charges}.

Using expressions (\ref{Koszul-explicit}) and (\ref{H2BPS}) one can explicitly check that indeed
\begin{equation}
  H_*(\CC,d)\cong \CH_\text{Top}
\label{HTop-iso}
\end{equation}  
as $\Z_2\times \Z^2$ graded vector spaces\footnote{The isomorphism should also be valid on the level of rings. The ring structure on $\CH_\text{Top}$ descends from the OPE structure on $\CH_\text{BPS}$. We will not perform this analysis here and leave it as an exercise.}. Let us write how the isomorphism map in (\ref{HTop-iso}) acts on the generators in (\ref{Koszul-explicit}). First,
\begin{equation}
 \Phi_1^{q_1} \longmapsto h_{q_1,0},\qquad q_1=0\ldots m-1.
\end{equation} 
The two subscripts of $h$ denote the charge with respect to $U(1)^{(1)}$ and $U(1)^{(2)}$ respectively and 
\begin{equation}
  h_{q_1,0} \in \left.\CH_{\alpha;\bar{\nu}}^{SU(2)_{mn-1}/U(1)_{2(mn-1)}}
	\otimes \CH_{(ma,n(a+\nu))}^{U(1)^2_Q} \right|_{\alpha=nq_1,a=0,\nu=-nq_1}.
\end{equation} 
is the primary from the first factor (coset module) tensored with the unique element from the second factor (lattice VOA module) that has $U(1)^2$ charges $(q_1,0)$ and minimal value of $L_0$. Such element exists because
\begin{equation}
 (q_1,0)= (0,-n^2q_1) \mod Q\Z^2.
\end{equation} 
The condition $L_0-\mathfrak{q}/2=0$ is satisfied because
\begin{equation}
\left.
 \frac{\alpha(\alpha+2)}{4(mn+1)}
-\frac{\nu^2}{4(mn-1)}
+\frac{1}{2}(\,q_1\;q_2\,)Q^{-1}\left(\begin{array}{c} q_1 \\ q_2\end{array}\right)-\frac{nq_1+mq_2}{2(mn+1)}
\right|_{\scriptsize\begin{array}{l} \alpha=nq_1,\\ \nu=-nq_1, \\ q_2=0\end{array}}=0.
\end{equation} 
Second, 
\begin{equation}
\bar\Psi_3\Phi_2^{n-1}\Phi_1^{q_1-1}-\bar\Psi_2\Phi_1^{q_1} \longmapsto h_{q_1,n},\qquad q_1=1\ldots m
\end{equation} 
where
\begin{equation}
  h_{q_1,n} \in \left.\CH_{\alpha;\bar{\nu}}^{SU(2)_{mn-1}/U(1)_{2(mn-1)}}
	\otimes \CH_{(ma,n(a+\nu))}^{U(1)^2_Q} \right|_{\alpha=nq_1-1,a=0,\nu=-nq_1+1}
\end{equation} 
is the primary from the first factor (coset module) tensored with the unique element from the second factor (lattice VOA module) that has $U(1)^2$ charges $(q_1,n)$ and minimal value of $L_0$. Again, such element exists because
\begin{equation}
 (q_1,n)= (0,n(-nq_1+1)) \mod Q\Z^2,
\end{equation} 
and the condition $L_0-\mathfrak{q}/2=0$ is satisfied since
\begin{equation}
\left.
 \frac{\alpha(\alpha+2)}{4(mn+1)}
-\frac{\nu^2}{4(mn-1)}
+\frac{1}{2}(\,q_1\;q_2\,)Q^{-1}\left(\begin{array}{c} q_1 \\ q_2\end{array}\right)-\frac{nq_1+mq_2}{2(mn+1)}
\right|_{\scriptsize\begin{array}{l} \alpha=nq_1-1,\\ \nu=-nq_1+1, \\ q_2=n\end{array}}=0.
\end{equation} 
The analysis for other elements in (\ref{Koszul-explicit}) is the same because of the symmetry under simultaneous exchange
\begin{equation}
 m\leftrightarrow n,\;q_1\leftrightarrow q_2,\; \Phi_1\leftrightarrow \Phi_2,\;
\Psi_1\leftrightarrow \Psi_2.
\end{equation} 

Finally let us note that the $U(1)^2$ equivariant Euler characteristic of $H_*(\CC,d)\cong \CH_\text{Top}$ can be computed from the superconformal index in the NS sector by rescaling flavor fugacities  and then taking $q\rightarrow 0$ limit \eqref{eulerchar}:
\begin{multline}
 	\left.\frac{\theta(q^{1-\frac{m+n}{2(mn+1)}}e^{-x-y})\theta(q^{1-\frac{mn}{2(mn+1)}}e^{-mx})\theta(q^{1-\frac{mn}{2(mn+1)}}e^{-ny})}
	{\theta(q^{\frac{n}{2(mn+1)}}e^{x})\theta(q^{\frac{m}{2(mn+1)}}e^y)}
\right|_{\scriptsize\begin{array}{l}e^x\rightarrow q^{-\frac{n}{2(mn+1)}}e^x
\\
\vspace{2em}
e^y\rightarrow q^{-\frac{m}{2(mn+1)}}e^y
\end{array}}\\
\stackrel{q\rightarrow 0}{\longrightarrow}\qquad
\frac{(1-e^{x+y})(1-e^{mx})(1-e^{ny})}{(1-e^x)(1-e^y)}.
\end{multline}

\subsection{$\cn=2$, $\fn=2$}
With two chiral fields and two Fermi fields we can write down the following Lagrangian 
\be
\int d \theta^+ \, \Psi_1 (\Phi_1^m+\Phi_2^n) +\Psi_2 \Phi_1\Phi_2,\qquad \qquad m,n\in {\mathbb Z}_+
\label{J-term-p2q2}
\ee
This model is simplest to analyze when either $m$ or $n$ is $1$, say $m=1$. The equation of motion for $\Psi_1$ implies $\Phi_1=-\Phi_2^n$. Integrating out $\Psi_1$ and $\Phi_1$, we get the interaction 
\be
\int d \theta^+ \, \Psi_2 \Phi_2^{n+1}.
\ee
This is the Lagrangian for $n$-th $\CN=(2,2)$ minimal model. 
Computation of the central charges and cohomology supports this conclusion. 
The trial right-moving central charge as a function of the R-charge of the Fermi field $\Psi_1$ is
\be
c_R=3\Big(\Big(1-\frac{1-r_{\Psi_1}}{n}\Big)^2+\Big(1-(1-r_{\Psi_1})\Big)^2-r_{\Psi_1}^2-\Big(1- \frac{1-r_{\Psi_1}}{n} - 1-r_{\Psi_1}\Big)^2\Big).
\ee
Extremizing with respect to $r_{\Psi_1}$,
\be
r_{\Psi_1}=\frac{2}{n+2}\qquad \Rightarrow \qquad c_R=\frac{3 \,n}{n+2}
\ee
This is exactly the central charge of the $n$-th minimal model. 
The gravitational anomaly is the difference between right-moving and left-moving central charge. It is given by
\be\label{gravan}
c_R-c_L={\rm Tr}\gamma^3.
\ee
In this model, the number of chiral and Fermi multiplet is the same and hence $c_L=c_R$. This is consistent with the $\CN=(2,2)$ minimal model.
We compute the topological heterotic ring of the Landau-Ginzburg model using \eqref{koszul}.
\bea
&& H_2(\CC,d)=0,\\
&& H_1(\CC,d)=0,\\
&& H_0(\CC,d)\cong {\rm Span}_{\mathbb C} \{ \Phi_1^i \}_{i=0}^{n}.
\eea
The only nontrivial cohomology is $H_0$ and it has dimension $n+1$ as expected of the $\CN=(2,2)$ minimal model.

For general values of $m$ and $n$, the c-extremization and vanishing of the gravitational anomaly  gives 
\be
c_L=c_R=\frac{3 \,mn}{mn+2}.
\ee
It is tempting to identify the infrared CFT again with the $\CN=(2,2)$ minimal model but computation of the cohomology rules out this possibility. For general values of $m,n$
\bea
&& H_2(\CC,d)=0,\\
&& H_1(\CC,d)=0,\\
&& H_0(\CC,d)\cong {\rm Span}_{\mathbb C} \{ \Phi_1^i \}_{i=0}^{n} \oplus {\rm Span}_{\mathbb C} \{ \Phi_2^j \}_{j=1}^{m-1}.
\label{p2q2-koszul}
\eea
The dimension of $H_0$ is $m+n$ and not $mn+1$ as expected from the $mn$-th $\CN=(2,2)$ minimal model. The right-moving supersymmetry guarantees that the right-moving part of the CFT is the chiral half of the $mn$-th $\CN=2$ minimal model but the left-moving part is not as straightforward to determine. We can still make progress by noting the existence of a left-moving $U(1)$ current. 
Under this symmetry the charges of the superfields $\Phi_{1,2}$ and $\Psi_{1,2}$ are $-m,\,-n$ and $mn,\, (m+n)$  respectively. Here we have chosen to normalize the charge so that they are integers. The 't Hooft anomaly for this symmetry is
\be
{\rm Tr}\gamma^3 \,FF=-(mn)^2-(m+n)^2 +m^2+n^2=-mn(mn+2).
\ee 
This abelian symmetry contributes $1$ to the left-moving central charge. The left-over central charge $c=2(mn-1)/(mn+2)<2$. As in the previous case we propose that it is contributed by chiral $SU(2)/U(1)$ WZW coset. 
The spectrum of the theory is then of the type
\be
\Big(\frac{SU(2)_{mn}}{U(1)_{2mn}} \times U(1)_{mn(mn+2)} \Big) \otimes \overline{\Big(\frac{SU(2)_{mn}\times SO(2)_1}{U(1)_{2(mn+2)}}\Big)}.
\label{p2q2-IR-CFT}
\ee
Where we have used again the coset representation of the $\CN=2$ minimal model on the right-moving side. The characters on the left-moving and right-moving side can be combined in a modular invariant way. This is anticipated by noticing that the level of $SU(2)$ piece is the same on both sides. As for the $U(1)$ factors, the existence of a modular invariant pairing follows from the rational equivalence of quadratic forms,
\be\label{Q-equiv}
\left(
\begin{array}{cc}
mn(mn+2) & 0\\
0 & 2(mn+2)
\end{array}
\right)
 \, \stackrel{\Q}{\sim} \,
 \left(
\begin{array}{cc}
2mn & 0\\
0 & 1
\end{array}
\right).
\ee
This follows from the same reasoning as in the previous subsection. The torus partition function is expressed in terms of affine characters as,
\begin{equation}
	Z=\sum_{\lambda,\alpha,\mu,\nu} C_{\alpha,\bar{\mu},\bar{\nu}}
	\,\chi_{\lambda;\bar{\alpha}}^{SU(2)_{mn}/U(1)_{2mn}}
	\chi_{\mu}^{U(1)_{mn(mn+2)}}\cdot 
	\bar{\chi}^{SU(2)_{mn}\times SO(2)_1/U(1)_{2(mn+2)}}_{\bar{\lambda};\nu}\,,
	\label{p2q2-Z}
\end{equation}
As the individual components of the low energy theory \eqref{p2q2-IR-CFT} depend only on the product $mn$, the separate dependence on $m$ and $n$ has to come from the coefficients $C_{\alpha,\bar{\mu},\bar{\nu}}$. As in the previous subsection, the choice of coefficients $C$ depends on the rational transformation that leads to the equivalence \eqref{Q-equiv}. Suppose the rational transformation,
\begin{equation}
	\left(\begin{array}{c}
	x \\
	y
	\end{array}
	\right)\;=\;
	\left(\begin{array}{cc}
	a & b \\
	c & d
	\end{array}\right)
	\left(\begin{array}{c}
	u \\
	v
	\end{array}
	\right),
	\qquad
	\left(\begin{array}{cc}
	a & b \\
	c & d
	\end{array}\right)\;\in\;GL(2,\Q)
	\label{p2q2-rat-trans}
\end{equation}
does the job \emph{i.e.},
\begin{equation}
	2mn\,x^2+y^2=mn(mn+2)\,u^2+2(mn+2)\,v^2.
	\label{p2q2-xy-uv}
\end{equation}

Although we could not find an analytic formula for the rational transform as function of a pair $(m,n)$, we checked in a number examples that for a given numbers $m$ and $n$ one can present an explicit rational transform which produces pairing coefficients $C_{\alpha,\bar{\mu},\bar{\nu}}$ for which the low energy calculation of the topological heterotic ring agrees with (\ref{p2q2-koszul}). We present a few cases below.

\subsubsection{Example: $(m,n)=(1,6)$}

The modular invariant pairing corresponding to the rational transform 
\begin{equation}
	\left(\begin{array}{c}
	x \\
	y
	\end{array}
	\right)\;=\;
	\left(\begin{array}{cc}
	1 & -1 \\
	6 & 2
	\end{array}\right)
	\left(\begin{array}{c}
	u \\
	v
	\end{array}
	\right).
\end{equation}
reads
\begin{equation}
	\CH=\bigoplus_{\lambda=0}^6\,
	\bigoplus_{\alpha\in\Z_{12}}\,
	\bigoplus_{s\in\Z_8}
	\CH_{\lambda;\bar{\alpha}}^{SU(2)_{6}/U(1)_{12}}
	\otimes
	\CH_{6s+\alpha}^{U(1)_{48}}
	\\
	\otimes 
	\bar{\CH}^{SU(2)_{6}\times SO(2)_1/U(1)_{16}}_{\bar{\lambda};2s-\alpha}.
\end{equation}
The BPS spectrum is given by
\begin{equation}
	\CH_\text{BPS}\equiv \CH|_{\bar{L}_0+\bar{J}_0/2=0}=
	\bigoplus_{\lambda=0}^6\,
		\bigoplus_{\ell\in\Z_{6}}
		\CH_{\lambda;2\ell+\lambda}^{SU(2)_{6}/U(1)_{12}}
		\otimes
		\CH_{8\ell+\lambda}^{U(1)_{48}}.
\end{equation}
The charge $\mathfrak{q}$ in (\ref{Top-bound}) which is used to define the topological heterotic ring is related to $q$, the charge of the $U(1)$ flavor symmetry as $\mathfrak{q}=q/(mn+2)=q/8$. It follows that the topological heterotic ring $\CH_\text{Top}$ is a 7-dimensional subspace of $\CH_\text{BPS}$: 
\begin{equation}
	\CH_\text{Top}\equiv \CH_\text{BPS}|_{L_0=\mathfrak{q}/2}=
	\mathrm{Span}_\C\{h_{\lambda,\lambda,\lambda}\}_{\lambda=0}^6
\end{equation}
where $h_{\lambda,\alpha,\mu}$ denotes the primary of $\CH_{\lambda;\alpha}^{SU(2)_{6}/U(1)_{12}}
		\otimes
		\CH_{\mu}^{U(1)_{48}}$. It is easy to see that this is in agreement with the Koszul homology (\ref{p2q2-koszul}).
		
		In fact, when either of $m$ or $n$ is $1$, say $m=1$, then the rational transformation 
\begin{equation}
	\left(\begin{array}{c}
	x \\
	y
	\end{array}
	\right)\;=\;
	\left(\begin{array}{cc}
	1 & -1 \\
	n & 2
	\end{array}\right)
	\left(\begin{array}{c}
	u \\
	v
	\end{array}
	\right).
\end{equation}
always gives the right low energy spectrum.
		
\subsubsection{Example: $(m,n)=(2,3)$}

One can choose the following rational transform:
\begin{equation}
	\left(\begin{array}{c}
	x \\
	y
	\end{array}
	\right)\;=\;
	\frac{1}{7}\,\left(\begin{array}{cc}
	11 & 5 \\
	30 & -22
	\end{array}\right)
	\left(\begin{array}{c}
	u \\
	v
	\end{array}
	\right).
\end{equation}
It produces the following modular invariant pairing
\begin{equation}
	\CH=\bigoplus_{\lambda=0}^6\,
	\bigoplus_{\alpha\in\Z_{12}}\,
	\bigoplus_{s\in\Z_8}
	\CH_{\lambda;\bar{\alpha}}^{SU(2)_{6}/U(1)_{12}}
	\otimes
	\CH_{6s+5\alpha}^{U(1)_{48}}
	\\
	\otimes 
	\bar{\CH}^{SU(2)_{6}\times SO(2)_1/U(1)_{16}}_{\bar{\lambda};2s-5\alpha}.
\end{equation}
The BPS spectrum is given by
\begin{equation}
	\CH_\text{BPS}\equiv \CH|_{\bar{L}_0+\bar{J}_0/2=0}=
	\bigoplus_{\lambda=0}^6\,
		\bigoplus_{\ell\in\Z_{6}}
		\CH_{\lambda;2\ell+\lambda}^{SU(2)_{6}/U(1)_{12}}
		\otimes
		\CH_{-8\ell+17\lambda}^{U(1)_{48}}.
\end{equation}
The topological heterotic ring $\CH_\text{Top}$ is its 5-dimensional subspace 
\begin{equation}
	\CH_\text{Top}\equiv \CH_\text{BPS}|_{L_0=\mathfrak{q}/2}=
	\mathrm{Span}_\C\{h_{0,0,0},h_{2,-2,2},h_{3,3,3},h_{4,-4,4},h_{6,6,6}\}
\end{equation}
Which is again in perfect agreement with the Koszul homology (\ref{p2q2-koszul}).

\subsubsection{Another example}
Finally, let us note that the rational transform
\begin{equation}
	\left(\begin{array}{c}
	x \\
	y
	\end{array}
	\right)\;=\;
	\left(\begin{array}{cc}
	2 & 0 \\
	0 & 4
	\end{array}\right)
	\left(\begin{array}{c}
	u \\
	v
	\end{array}
	\right).
\end{equation}
also implies (\ref{p2q2-xy-uv}) with $mn=6$. It provides the following modular invariant pairing:
\begin{equation}
	\CH=\bigoplus_{\lambda=0}^6\,
	\bigoplus_{\alpha\in\Z_{12}}\,
	\bigoplus_{s\in\Z_2}
	\bigoplus_{r\in\Z_4}
	\CH_{\lambda;\bar{\alpha}}^{SU(2)_{6}/U(1)_{12}}
	\otimes
	\CH_{24s+2\alpha}^{U(1)_{48}}
	\\
	\otimes 
	\bar{\CH}^{SU(2)_{6}\times SO(2)_1/U(1)_{16}}_{\bar{\lambda};4r}.
	\label{pairing-non-LG}
\end{equation}
However the resulting topological heterotic ring $\CH_\text{top}$ is 1-dimensional and therefore does not correspond to a LG model (\ref{J-term-p2q2}) for any $m$ and $n$. The sums over $\alpha$ and $s$ in can be performed and give $SU(2)_6$ modules in the left-moving sector:
\begin{equation}
	\CH=\bigoplus_{\lambda=0}^6\,
	\bigoplus_{r\in\Z_4}
	\CH_{\lambda}^{SU(2)_{6}}
	\\
	\otimes 
	\bar{\CH}^{SU(2)_{6}\times SO(2)_1/U(1)_{16}}_{\bar{\lambda};4r}.
\end{equation}
This is a ``$(0,2)$ minimal model'' of the type considered in \cite{Berglund:1995dv}. In general, when $mn=2(Q^2-1),\;Q\in\Z$, there is a rational transform
\begin{equation}
		\left(\begin{array}{c}
		x \\
		y
		\end{array}
		\right)\;=\;
		\left(\begin{array}{cc}
		Q & 0 \\
		0 & 2Q
		\end{array}\right)
		\left(\begin{array}{c}
		u \\
		v
		\end{array}
		\right)
\end{equation}
which produces modular invariant pairing that does not correspond to a LG model (\ref{J-term-p2q2}), but corresponds to a model in \cite{Berglund:1995dv}.

\section{Solvable models with non-abelian symmetry}
The Landau-Ginzburg models studied so far did not have any non-abelian symmetries. In this section, we consider a model that has a $U(N)$ symmetry. At low energy this symmetry enhances to an infinite dimensional affine symmetry. 
Using the arguments of 't Hooft anomaly matching and Sugawara central charge saturation we are able to determine its low energy theory. The philosophy is similar to that of \cite{Gadde:2014ppa}.

Consider a family of $(0,2)$ LG theories labeled by $N\in\Z_+$ that have $U(N)=(SU(N)\times U(1))/\Z_N$ flavor symmetry and the following matter content:
\begin{itemize}
 \item Chiral multiplet in anti-fundamental ($\afund$) representation: $\Phi_i$
 \item Fermi multiplet in symmetric ($\sym$) representation: $\Psi^{ij}$
\end{itemize}
and the $J$-type superpotential
\begin{equation}
 \int d\theta^+\,\Psi^{ij}\Phi_i\Phi_j\;\;+\;\;\text{c. c.}
\end{equation} 
Note that when $N=2$ the supersymmetry is enhanced to (2,2) and the theory at hand flows to the first non-trivial $N=2$ minimal model in both left- and right-moving sectors \cite{Witten:1993jg}. The theory has the following 't Hooft and gravitational anomalies:
\begin{equation}
 \Tr \gamma^3 J_{SU(N)}^2=T(\sym)-T(\afund)=\frac{N+1}{2}
\end{equation} 
\begin{equation}
 \Tr \gamma^3 J_{U(1)}^2=4\frac{N(N+1)}{2}-N=2N^2+N
\end{equation} 
\begin{equation}
 c_\text{L}-c_\text{R}=\Tr \gamma^3 =\frac{N(N+1)}{2}-N=\frac{N(N-1)}{2}
\end{equation} 
Since 't Hooft anomalies are protected under RG flow it follows that the left-moving sector of the CFT in the infrared contains the following WZW chiral algebra:
\begin{equation}
 \text{CFT}_\text{L}\;\supset\; SU(N)_{N+1}\times U(1)_{N(2N+1)} \label{WZW1}
\end{equation} 
The theory has a normalizable vacuum state (classically all bosonic fields vanish) and so one can apply $c$-extremization procedure \cite{Benini:2013cda} to the following probe right-moving central charge:
\begin{equation}
 c_\text{R}=3\,\Tr\gamma^3\,R^2=3\,N\,(R_\Phi-1)^2-3\,\frac{N(N+1)}{2}\,R_\Psi^2
\end{equation} 
where $R_\Phi$ and $R_\Psi$ are R-charges of the chiral and Fermi multiplets respectively which are subject to the following constraint given by the $J$-type superpotential:
\begin{equation}
 2R_\Psi+R_\Phi=1
\end{equation} 
The result gives us the following IR values:
\begin{equation}
 R_\Phi=\frac{N}{2N+1}\qquad R_\Psi=\frac{1}{2N+1},
\end{equation} 
\begin{equation}
 c_\text{R}=\frac{3N(N+1)}{4N+2}\qquad c_\text{L}=\frac{N(N^2+N+1)}{2N+1}
 \label{cIR}
\end{equation} 
Interestingly, the left-moving central charge coincides with the Sugawara central charge of WZW model in (\ref{WZW1}). In general from (\ref{WZW1}) one only expects inequality $c_\text{L}\geq c_\text{Sugawara}$. Since it is saturated, it follows that in the left moving sector of the IR CFT should coincide with chiral WZW in (\ref{WZW1}):
\begin{equation}
 \text{CFT}_\text{L}=SU(N)_{N+1}\times U(1)_{N(2N+1)} 
 \label{CFTL}
\end{equation} 

We conjecture that the right-moving sector is described in terms of a KS coset model with $\CN=2$ supersymmetry \cite{KS1,KS2}:
\begin{equation}
 \text{CFT}_\text{R}=\left[\frac{SO(2N+2)}{U(N+1)}\right]_{2N+1}\cong
\frac{SO(2N+2)_1\times SO(N(N+1))_1}{SU(N+1)_N\times U(1)_{(N+1)(2N+1)}}
\label{CFTR}
\end{equation} 
Where the second equality expresses the supersymmetric WZW coset as an ordinary bosonic coset. Note that $N(N+1)$ is the difference between dimensions of $SO(2N+2)$ and $U(N+1)$. The claim is based on the fact that the central charge is indeed as in (\ref{cIR}) and, most importantly, on the fact that there is a natural pairing between primaries of (\ref{CFTL}) and (\ref{CFTR}) that gives a modular invariant\footnote{By modular invariance we mean invariance up to an overall simple factor due to 't Hooft and gravitational anomalies. Also, for the partition function in the NS-NS sector, the modular group which has to be considered is not the full $SL(2,\Z)$ but its subgroup generated by $S$ and $T^2$ elements.} partition function. The argument is parallel to the one made in \cite{Gadde:2014ppa} where we refer the reader for details. Below we present the main ideas. The pairing is a version of level-rank duality and is given explicitly by the following conformal embedding:
\begin{multline}
	U(1)_1\times U(N(N+1))_1 \;\cong \\ \left(SU(N)_{N+1}\times U(1)_{N(2N+1)}\right)
	\times  \left(SU(N+1)_{N}\times U(1)_{(N+1)(2N+1)}\right)
\end{multline}
Since the chiral algebra in the left hand side has only one irreducible module, characters of the two factors in the right hand side transform in conjugate representations of the modular group. The numerator in the right hand side of (\ref{CFTR}) has a natural module invariant under the action of modular group which can be realized by free chiral fermions. 

\acknowledgments{The authors gratefully acknowledges support from the Institute for Advanced Study. AG is supported by the Roger Dashen Membership Fund and the National Science Foundation grant PHY-1314311.
This work was performed in part (by PP) at Aspen Center for Physics which is supported by National Science Foundation grant PHY-1066293. 
}


\bibliographystyle{JHEP_TD}
\bibliography{02-refs}

\end{document}